\documentclass[aps,prl,twocolumn,showpacs,superscriptaddress,preprintnumbers,amsmath,amssymb]{revtex4-1}

\usepackage{graphicx}
\usepackage{epsfig}
\usepackage{dcolumn}
\usepackage{bm}
\usepackage{longtable}
\usepackage{color}

\begin{document}


\title{Bulk evidence for a time reversal symmetry broken superconducting state in URu$_2$Si$_2$}



\author{G.\ Li}
\affiliation{National High Magnetic Field Laboratory, Florida
State University, Tallahassee-FL 32310, USA}
\author{Q.\ Zhang}
\affiliation{National High Magnetic Field Laboratory, Florida
State University, Tallahassee-FL 32310, USA}
\author{D.\ Rhodes}
\affiliation{National High Magnetic Field Laboratory, Florida
State University, Tallahassee-FL 32310, USA}
\author{B.\ Zeng}
\affiliation{National High Magnetic Field Laboratory, Florida
State University, Tallahassee-FL 32310, USA}
\author{P.\ Goswami}
\affiliation{National High Magnetic Field Laboratory, Florida
State University, Tallahassee-FL 32310, USA}
\author{R. E. Baumbach}
\affiliation{Los Alamos National Laboratory, Los Alamos, New Mexico 87545, USA}
\author{ P.\ H.\ Tobash}
\affiliation{Los Alamos National Laboratory, Los Alamos, New Mexico 87545, USA}
\author{F.\ Ronning}
\affiliation{Los Alamos National Laboratory, Los Alamos, New Mexico 87545, USA}
\author{J.\ D.\ Thompson}
\affiliation{Los Alamos National Laboratory, Los Alamos, New Mexico 87545, USA}
\author{E.\ D.\ Bauer}
\affiliation{Los Alamos National Laboratory, Los Alamos, New Mexico 87545, USA}
\author{L.\ Balicas} \email{balicas@magnet.fsu.edu}
\affiliation{National High Magnetic Field Laboratory, Florida
State University, Tallahassee-FL 32310, USA}


\date{\today}

\begin{abstract}
\textbf{URu$_2$Si$_2$ is claimed to be a chiral \emph{d}-wave superconductor with a $k_z (k_x \pm  ik_y)$ time-reversal symmetry broken orbital component for the Cooper pair wave-function, which contains both nodal points and lines of nodes \cite{kasahara, kasahara2}. To study the magnetic response of such an unconventional state through a bulk, thermodynamic probe, we measured the magnetic torque $\tau$ in very high-quality, well-characterized URu$_2$Si$_2$ single-crystals \cite{altarawneh,altarawneh2} at high magnetic-fields $H$ and at very low temperatures $T$.  The magnetization $M(H) \propto \tau(H) /H$ of URu$_2$Si$_2$, in its superconducting state and for angles within $15^{\circ}$ from the \emph{ab}-plane, reveals a change in its sign for $H$ approaching $H_{c2}$: from a clear diamagnetic response dominated by the pinning of vortices to a state with a smaller but ``paramagnetic-like" hysteretic response which \emph{disappears} at $H_{c2}$, thus implying that it is intrinsically related to the superconducting state.  We argue that this anomalous, angular-dependent behavior is evidence for a time-reversal symmetry broken superconducting state in URu$_2$Si$_2$, although not necessarily for the $k_z (k_x \pm  ik_y)$ state.}

\end{abstract}


\maketitle

The nature of the hidden-order (HO) state in URu$_2$Si$_2$ and its interplay with superconducting and antiferromagnetic states continues to be the subject of intense scrutiny.
The observation of Fano-like resonances in the quasi-particle interference patterns as measured through scanning tunneling spectroscopy \cite{fano1} or through point contact spectroscopy \cite{fano2}, confirms the development
of a Kondo-lattice (a lattice of composite quasi-particles resulting from the hybridization between localized $f$-moments and itinerant $d$ carriers) in the metallic state preceding the HO-phase at $T_{\text{HO}} \simeq 17.5$ K.

Most theoretical proposals for the HO-state fall into two categories: the first analyzes $k$-space susceptibilities at the Fermi surface ascribing the HO to density-wave like phases \cite{maple,balatsky,ikeda,varma,chandra} while the second one considers the local ordering in real-space of states at the U sites, with the corresponding alteration (via changes in the hybridization between itinerant and localized states) to the band structure \cite{haule,broholm2,barzykin,santini,haule2,harima}. To date, proposals in neither category have been unambiguously proven to accurately describe the transition towards the HO-state. Although a recent theoretical proposal \cite{ikeda2} claims that all of the above properties can be reconciled with a rank-5 multipole, i.e. a dotriacontapole order-parameter having $E^{-}$ nematic symmetry and exhibiting staggered pseudospin moments along the [110] direction. A very recent analysis of the magnetic field-induced magnetization distribution around the U ions by polarized neutron elastic-scattering measurements, claims to support this scenario \cite{ressouche}.

Since in URu$_2$Si$_2$ an unconventional metallic state, e.g. characterized by strong spin fluctuations \cite{wiebe}, and an exotic ordered-state close to magnetism precedes superconductivity, it is natural to expect an unconventional superconducting state for this material. In effect, the temperature dependence of both the specific heat, i.e. $C(T)\propto T^2$, \cite{hasselbach} and of the nuclear magnetic resonance relaxation rate,  $T_{1}^{-1} \propto  T^3$ \cite{kohori} at low-temperatures are indications for a line node in the superconducting gap or for a density of states (DOS) $N(E) \propto |E|$ at low energies. Transport experiments indicates that the electronic structure of the HO-state \cite{altarawneh,kasahara} is composed of both electron and hole states thus indicating that URu$_2$Si$_2$ is a multiband superconductor. Thermal conductivity $\kappa$ \cite{kasahara, kasahara2} reveals: i) an electronic contribution in the limit of very low temperatures which can be attributed to the presence of nodes in the superconducting gap function, ii) at low magnetic fields $\kappa$ is proportional to $\sqrt{H}$ indicating the presence of node lines, iii) at higher fields $\kappa$ becomes strongly and anomalously field-dependent displaying a sharp step-like reduction at the upper critical-field(s) $H_{c2}$, interpreted as an indication for a \emph{first-order} phase-transition at $H_{c2}$ (for $T \lesssim 500$ mK) and iv) the angular dependence of the electronic contribution to $\kappa$ was claimed to be consistent with the existence of two distinct superconducting gaps, in which horizontal line-nodes would lie within the basal \emph{ab}-plane of a light hole-like band having a small superconducting gap and point nodes along the \emph{c}-axis in a heavy, electron-like band having a large gap. These observations were claimed to be consistent with a chiral, time-reversal symmetry-breaking two-component order- parameter having \emph{d}-wave symmetry with the form $\Delta_k \propto k_z (k_x \pm  ik_y)$ \cite{kasahara, kasahara2}. Finally, the orientation dependence of $H_{c2}$ indicates a very anisotropic effective Land\'{e} g-factor (as estimated from the Pauli limiting field) implying an extremely anisotropic spin susceptibility \cite{altarawneh2}. This suggests that the quasiparticles subject to pairing in URu$_2$Si$_2$ might be ``composite heavy fermions" formed from bound states between the conduction electrons and local \emph{f}-moments which have a protected Ising-like behavior.

The expression $k_z (k_x \pm  ik_y)$ for the pair wave-function corresponds to an even orbital-function that breaks time-reversal symmetry (TRS) because the superconducting condensate acquires an overall orbital magnetic moment. While a perfect sample might not exhibit a net moment, in principle, surfaces and defects at which the Meissner screening of the TRS-breaking moment is not perfect can lead to a small magnetic signal \cite{manfred}. Since the existence of TRS-breaking has considerable implications for the superconducting state of URu$_2$Si$_2$, observing this effect, particularly in the bulk, without relying on imperfections or defects is of the utmost importance. The experimental challenge is to couple to the TRS-breaking part of the order-parameter to demonstrate this effect unambiguously. Here, we present torque magnetometry measurements at very low temperatures and in a high quality single-crystal of URu$_2$Si$_2$ at high magnetic fields with the goal of detecting evidence for TRS-breaking in its superconducting state.

\begin{figure}[htbp]
\begin{center}
\epsfig{file=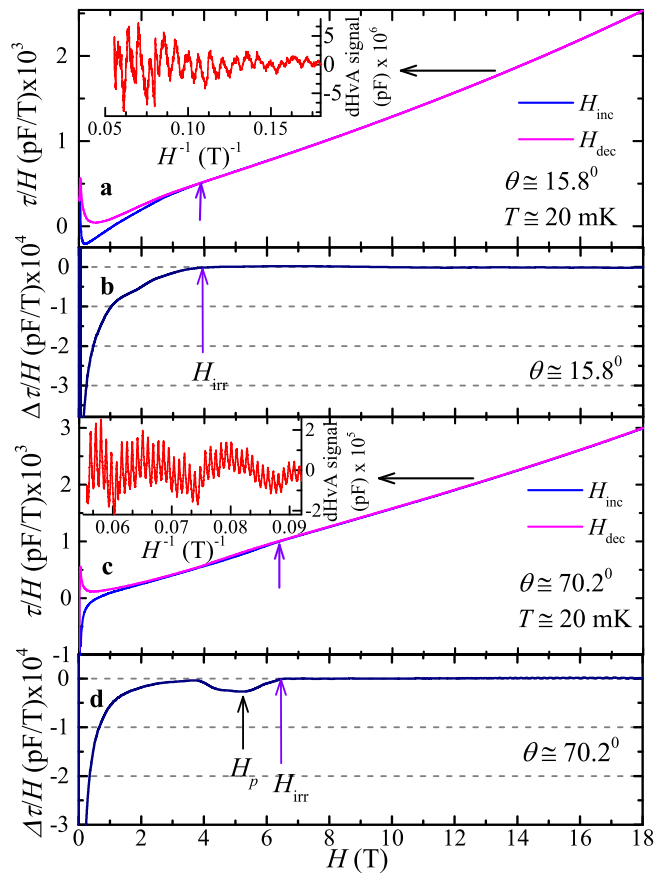, width = 8.6 cm}
\caption{\textbf{Field and angular dependence of the magnetic torque in URu$_2$Si$_2$.} \textbf{a} Magnetic torque $\tau$ normalized by the external magnetic field $H$ as a function of $H$ for a URu$_2$Si$_2$ single-crystal at a temperature $T = 20 $ mK. The angle $\theta$ between the c-axis and the magnetic field is $\theta =15.8^{\circ}$. Blue and magenta lines correspond to field-up and down sweeps respectively, purple arrow indicates the irreversibility field.  \textbf{b} Irreversible/hysteretic component in the magnetic torque, $\Delta \tau /H$ from the traces in \textbf{a}. \textbf{c} Same as in \textbf{a} but for an angle $\theta = 70.2^{\circ}$. \textbf{d} Same as in \textbf{b} but for $\theta = 70.2^{\circ}$. Black arrow indicates the field $H_{p}$ where a minimum is observed in $\Delta \tau /H$. Insets: oscillatory component, i.e. de Haas van Alphen-effect, superimposed into $\tau /H$.}
\end{center}
\end{figure}
\begin{figure}[htb]
\begin{center}
\epsfig{file=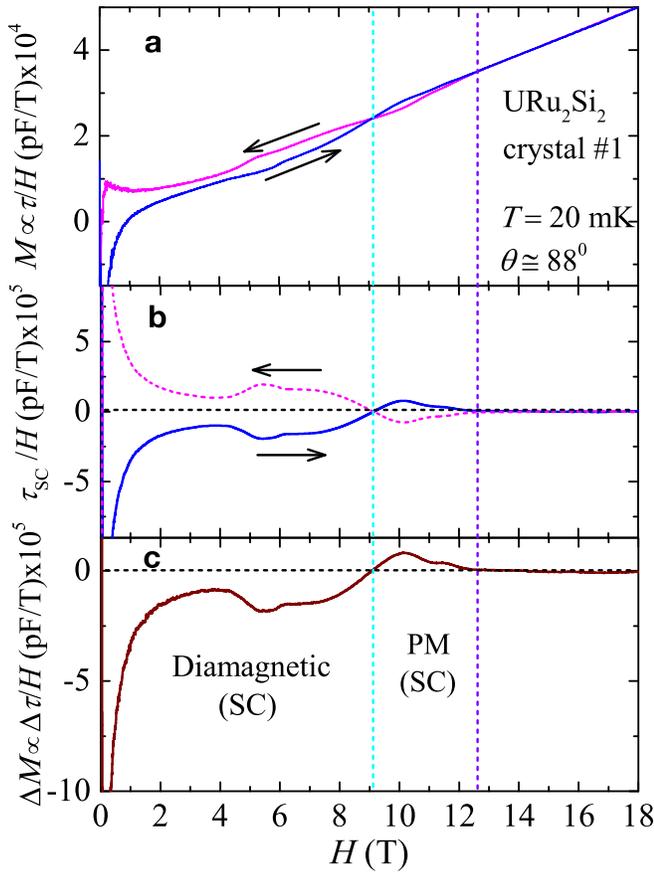, width = 8.6 cm}
\caption{\textbf{Emergence of a paramagnetic response \emph{within} the superconducting state of URu$_2$Si$_2$.}\textbf{a} $ M \propto \tau/H$ (where $M$ is the magnetization) as a function of the field $H$ for an angle $\theta = 88^{\circ}$ and at a temperature $T = 20$ mK. Again, blue and magenta lines depict field-increasing and decreasing sweeps. As expected, or above $H_{c2}$ (indicated by the purple vertical line) $\tau/H$ is linear in $H$, which is the behavior expected for an anisotropic paramagnetic metallic state. \textbf{b} $\tau_{\text{SC}}$, or the difference between the raw $\tau/H$ traces shown in \textbf{a} and their average. This simple procedure subtracts the strong metallic paramagnetic background. Notice, how the remanent non-linear response (e.g. blue line), due solely to the superconducting state, crosses zero thus indicating that the magnetic response of the superconducting state just below $H_{c2}$ is paramagnetic (PM) in nature and not diamagnetic. \textbf{c} $\Delta \tau/H$ or the difference in $\tau/H$ for the field-up and -down sweeps shown in \textbf{a}, and which corresponds to the pure hysteretic response (brown line). Notice how the observed anomalous hysteresis is associated with the paramagnetic like response in $M \propto \tau/H $. Vertical cyan line indicates the field where the anomalous hysteresis emerges.}
\end{center}
\end{figure}

Figure 1\textbf{a} displays the magnetic torque $\overrightarrow{\tau} = \mu_0 \overrightarrow{M} \times \overrightarrow{H}$ ($\overrightarrow{M}$ is the magnetization of the sample) for a URu$_2$Si$_2$ single crystal as a function of the field $H$, normalized by $H$ at a temperature $T \simeq 20$ mK and at an angle $\theta = 15.8^{\circ}$ between the external field and the inter-plane \emph{c}-axis. Blue and magenta lines indicate field-increasing $H_\text{inc}$ and decreasing $H_\text{dec}$ sweeps, respectively. For a layered system, and assuming a uniform in plane-susceptibility $\chi_{aa}$, one can readily demonstrate
that $\tau = \mu_0/2 (\chi_{aa} - \chi_{cc})H^2 \sin 2 \theta$  where  $\chi_{cc}$ is the inter-planar component of the susceptibility tensor. Thus, the torque is extremely sensitive to the magnetic response of magnetically anisotropic systems. As mentioned above and as discussed in Refs. \onlinecite{altarawneh2,chandra}, URu$_2$Si$_2$ in its hidden-order state is extremely anisotropic which leads to a very large, nearly linear $\tau(H)/H$ which becomes progressively smaller as one approaches a crystallographic axis of symmetry, such as the \emph{ab}-plane. The hysteretic response in $\tau(H)/H$, shown in Fig. 1\textbf{b} corresponds to the diamagnetic hysteretic component, defined as the difference between both branches $ \Delta \tau/H = (\tau(H_\text{inc})/H_\text{inc} - \tau(H_\text{dec})/H_\text{dec})$ which according to the Bean model \cite{bean} is proportional to the critical current density (and concomitant vortex pinning force(s)).
\begin{figure}[htb]
\begin{center}
\epsfig{file=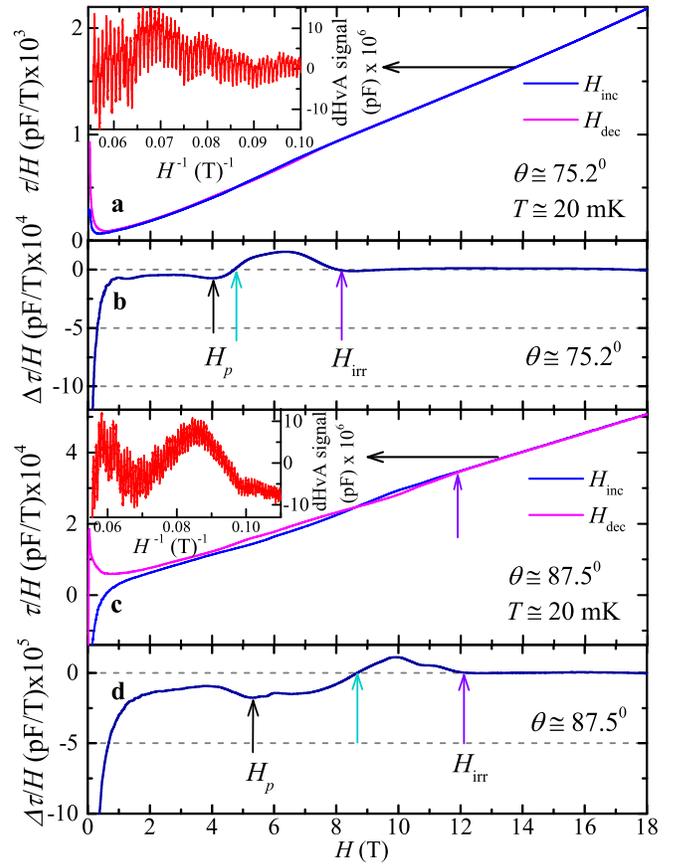, width = 8.6 cm}
\caption{\textbf{Angular dependence of the $\tau$ for fields close to the \emph{ab}-plane}. \textbf{a} Magnetic torque $\tau$ normalized by the external magnetic field $H$ as a function of $H$ for a URu$_2$Si$_2$ single-crystal at a temperature $T = 20 $ mK. The angle $\theta$ between the c-axis and the magnetic field is $\theta =75.2^{\circ}$. Blue and magenta lines correspond to field-up and down sweeps respectively, purple arrow indicates the irreversibility field. Black arrow indicates $H_{p}$, clear blue arrow indicates the value in field where $\Delta \tau /H$ crossovers from negative to positive values. \textbf{b} Irreversible or hysteretic component in the magnetic torque, $\Delta \tau /H$ from the traces in \textbf{a}. \textbf{c} Same as in \textbf{a} but for an angle $\theta = 87.5^{\circ}$. \textbf{d} Same as in \textbf{b} but for $\theta = 87.5^{\circ}$. Insets: oscillatory component superimposed into $\tau /H$.}
\end{center}
\end{figure}
\begin{figure}[htb]
\begin{center}
\epsfig{file=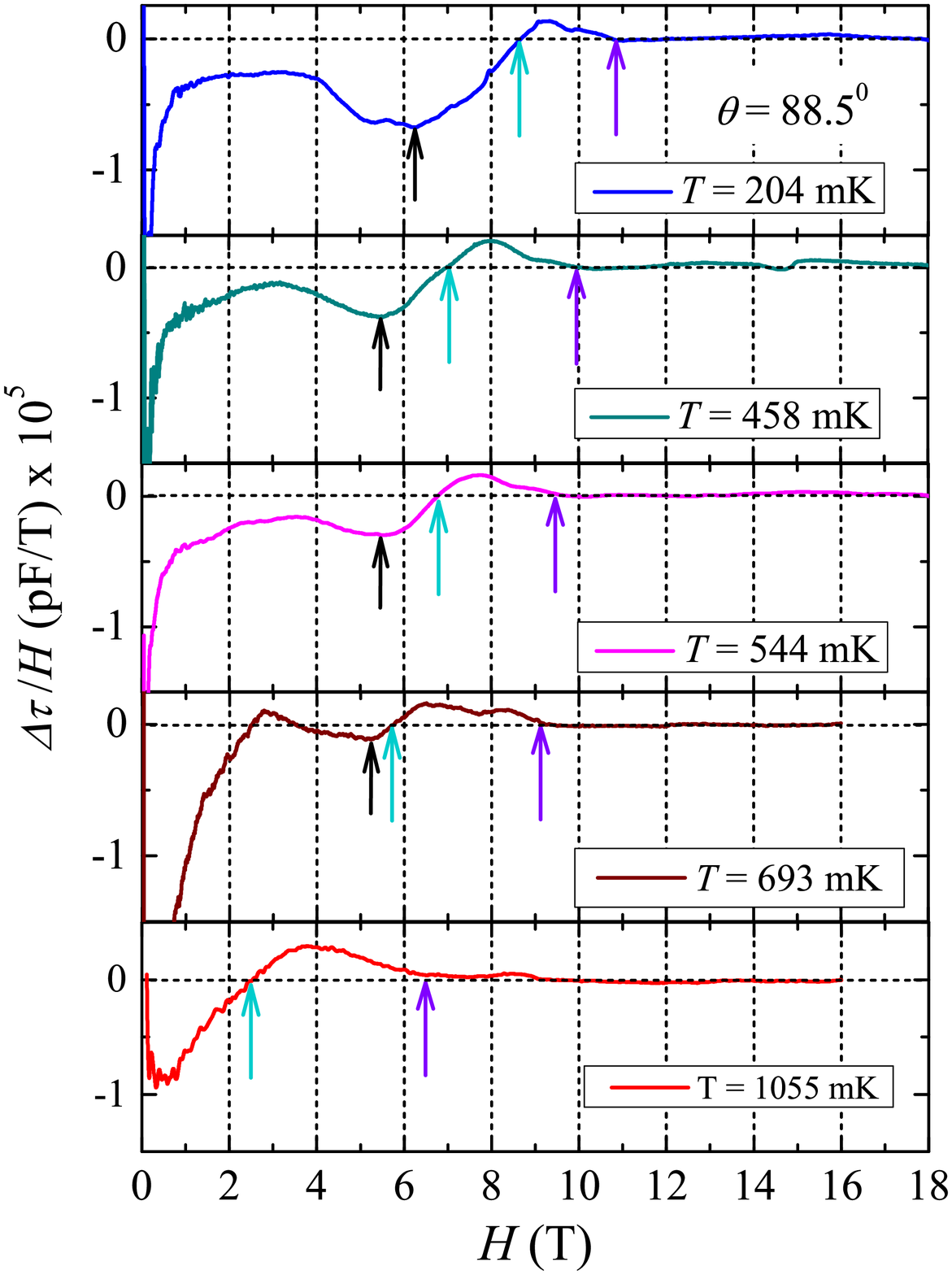, width = 8.6 cm}
\caption{\textbf{Temperature-dependence of the anomalous superconducting response}. $ \Delta \tau/H $ as a function of the field for $\theta=88.5^{\circ}$ and for several temperatures. Black arrow indicates the minimum at $H_{\text{dia}}$, purple arrow the irreversibility field, and clear blue arrow the crossover from diamagnetic to paramagnetic-like hysteresis.}
\end{center}
\end{figure}
Notice, how the hysteresis is small when compared to the linear paramagnetic (PM) background, suggesting small vortex pinning forces. This is consistent with the extreme high purity of this crystal, i.e. the presence of very few vortex point pinning centers. It is most pronounced at very low fields, i.e. $H \lesssim 2$ T, or in the field region where the thermal conductivity is observed to behave as $\sqrt{H}$ \cite{kasahara}.
For $\theta = 15.8^{\circ}$ one sees conventional hysteretic behavior that progressively disappears as $H$ increases. But for $\theta = 70.2^{\circ}$,
$ \Delta \tau/H$ behaves in an unexpected way, displaying almost no hysteresis for $H \sim 4$ T but a remarkable enhancement in the hysteretic response above this value. This suggests a transition from a state akin to a vortex liquid, or characterized by the absence of vortex pinning, to a state with an enhanced pinning response (i.e. a vortex solid) similar to the so-called fishtail or peak-effect which is still poorly understood and
is frequently explained as i) a transition from a vortex-ordered to a -disordered state \cite{ordertodisorder}, or ii) to a competition between surface barriers and bulk pinning \cite{surfacetobulk}.
This enhancement is observed for $\theta \gtrsim 30^{\circ}$, so here thereafter we name the region $\theta \leqslant 30^{\circ}$ as region I, and region II as the angular window characterized by this anomalous enhancement in pinning/hysteresis, i.e. $ 30^{\circ} \leqslant \theta \leqslant 75^{\circ}$ (see text below).

In figure 2, we subtract from $M \propto \tau/H$, at $T = 20$ mK and at a fixed angle $\theta = 88^{\circ}$ with respect to the $c-$axis, the paramagnetic metallic contribution. As previously seen, for both field-up (blue line) and -down (magenta line) sweeps, the hysteresis due solely to the superconducting response, is small when compared to the nearly linear paramagnetic background. In the superconducting state this background results from the contribution of the metallic vortex-cores whose density per unit area increases as $H$ increases. Already for fields as small as $\sim 3$ T this background becomes comparable in size to the hysteretic superconducting contribution. Therefore, if one takes the average between field-up and -down branches one obtains an average $\tau_{\text{av}}/H$ curve, which contains contributions from both the superconducting and the paramagnetic metallic states, but mostly from this last one at high fields. Consequently, if one subtracts $\tau_{\text{av}}/H$ from each $\tau/H$ branch, one obtains a magnetic response $M_{\text{SC}} \propto \tau_{\text{SC}}/H$ which no longer contains the paramagnetic metallic contribution, but solely the magnetic response from the superconducting state. Remarkably, and this is the main observation in this manuscript, as indicated by the blue line in Fig. 2 \textbf{b}, the magnetic response due solely to superconductivity, is observed to cross the zero value, indicating that it crosses-over (at certain field value indicated by the cyan vertical line) from a net diamagnetic response due to vortex pinning, to a net paramagnetic-like but still hysteretic response. This paramagnetic-like response disappears at the irreversibility field $H_{\text{irr}}$ (the higher value in field where the increasing and decreasing $M_{\Delta}$ branches meet and reach a value of zero, as indicated by the violet vertical line) thus clearly implying that it is intrinsic to the superconducting state. This unexpected response also leads to an anomalous net hysteresis in $ \Delta \tau/H$ as seen in Fig. 2 \textbf{c}, where the net diamagnetic hysteresis due to pinning is followed by an anomalous hysteretic response of opposite sign. In the remainder of the text we will follow the angular- and the temperature-dependence of this abnormal paramagnetic response in the magnetization, through this anomalous hysteretic behavior.
\begin{figure}[htb]
\begin{center}
\epsfig{file=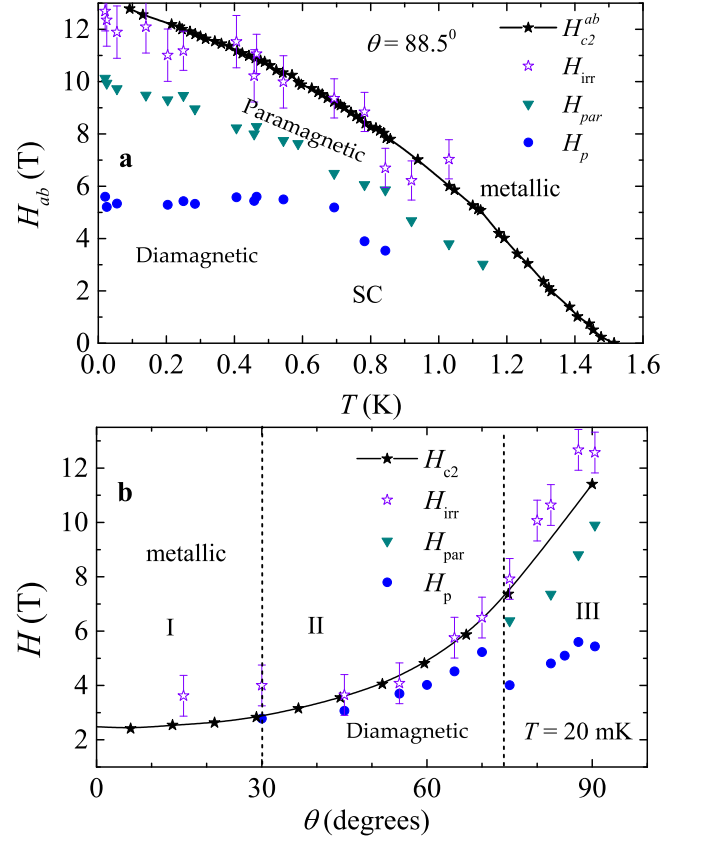, width = 8.6 cm}
\caption{ \textbf{Temperature and angle dependent phase-diagrams}. \textbf{a} Resulting $H$ as a function of $T$ phase diagram for an angle $\theta \simeq 88.5^{\circ}$. Black markers correspond
to the phase boundary from Ref. \onlinecite{brison}, violet ones to the irreversibility field $H_{\text{irr}}$, clear blue markers to the field $H_{\text{par}}$ where the hysteretic response changes sign, and blue markers to the field $H_p$ corresponding to the maximum in the enhanced diamagnetic response. \textbf{b} $H$ as a function of $\theta$ phase-diagram for $T = 20$ mK showing zones I, II, and III, respectively. Black markers depict the boundary from Ref. \onlinecite{altarawneh2}. Notice that the boundary defined by $H_p$ splits into two when going from zones II to III.}
\end{center}
\end{figure}

One could argue that this anomalous response in $\tau/H$ could result from a change in the magnetic anisotropy of the superconducting state, e.g. from vortices piercing the superconducting planes to vortices pinned in-between the planes (i.e. intrinsic-pinning). Of course, the large anisotropy of URu$_2$Si$_2$ would prevent such a scenario, in addition to the fact that  $H_{c2}^{ab} \simeq 12.3$ T corresponds to an inter-planar coherence length $\xi_c \simeq 23 $ \AA  (for $\xi_{ab} = 114.8$ \AA) which is larger than the inter-planar distance \cite{palstra} $c = 9.5817(8)$ \AA. Therefore, one cannot conceive a plausible physical scenario where a simple vortex re-configuration could possibly lead to a change in the relative weight between $\chi_{aa}$ and $\chi_{zz}$ leading the anomalous PM response seen by us. Unless of course, the vortex cores themselves developed a stronger PM signal than the bare metallic state seen at higher fields. Finally, we show in the Supplemental Information section, that this anomalous behavior is \emph{reproducible} and observed \emph{only} in the highest quality single-crystals.

Figure 3 shows both $\tau/H$ and $ \Delta \tau(H)$ as a function of $H$ at a temperature of 20 mK and for two other values of $\theta$,  respectively $75.2^{\circ}$ and $87.5^{\circ}$. As previously seen, for both angles the hysteresis is most pronounced for $H \lesssim 1$ T. One observes, as in Fig. 1\textbf{d}, an enhancement in the diamagnetic response for fields in the neighborhood of $H=4$ T, which in both cases is followed by an anomalous change, from negative to positive in the sign of both the torque signal (as seen in Fig. 2) and in the hysteretic response (indicated by the clear blue arrow), associated with the \emph{paramagnetic-} or (\emph{magnetic-}) response \emph{within the superconducting state} as $ H \rightarrow H_{c2}$. Here, we must emphasize that any extrinsic (or intrinsic) magnetic signal superimposed onto the superconducting one \emph{cannot} reproduce the anomalous hysteresis: e.g. both ferromagnetism and superconductivity leads to the exact same sign for the net hysteretic response between field-up and -down sweeps. This anomalous paramagnetic superconducting response and concomitant anomalous hysteresis is detected only within the angular window $ 75^{\circ} \leqslant \theta \leqslant 90^{\circ}$ or in region III.

Figure 4 displays $\Delta \tau/H$ as a function of $H$ for several temperatures (indicated in the figure) and for $\theta = 88.5^{\circ}$. As seen the anomalous superconducting paramagnetic response and related hysteresis moves to lower fields as $T$ increases tracking the behavior of $H_{\text{irr}} \simeq H_{c2}(T)$ \cite{brison} and thus further confirming that it is intrinsically related to the superconducting state.

From the curves in Fig. 3 we built the $H$ as a function of $T$ phase-diagram displayed in Fig. 5\textbf{a} for an angle $\theta = 88.5^{\circ}$.
As seen, the anomalous hysteresis follows the phase-boundary between metallic and superconducting states up to higher $T$s, while
the boundary defining the enhanced diamagnetic response is nearly $T$-independent up to $T=0.7$ K, decreases in field beyond this value, and becomes undetectable for $T \gtrsim 0.83$ K. In Fig 5\textbf{b} we show the $H$ as a function of $\theta$ phase-diagram, indicating all three zones. In zone III, $H_p$ shifts to lower fields as it is displaced by the emergence of the paramagnetic response at $\theta \gtrsim 75^{\circ}$.

A paramagnetic-like \emph{Meissner} response seen \emph{at very low fields} and known as the Wohlleben-effect, has been reported in a variety of superconducting systems \cite{geim,moshchalkov,koshelev,sigrist}. Most explanations for the Wohlleben-effect fall into two categories: (i) finite system size- and inhomogeneous cooling- induced surface superconductivity, which causes flux compression for fields below the first upper-critical field \cite{koshelev}; and (ii) granularity-induced random $\pi$ junctions in $d$-wave superconductors \cite{sigrist}. In the above two mechanisms, the paramagnetic response becomes larger than the diamagnetic response only in the low-field ($H<H_{c1}$) \emph{Meissner} state. But in our experiments, the paramagnetic response dominates the diamagnetic one only in the very high-field mixed state (i.e. close to $H_{c2}$), or when the field has penetrated the surface of the crystal through a length beyond the penetration depth, which is inconsistent with the Wohlleben-effect and associated models. Furthermore, this anomalous hysteresis in URu$_2$Si$_2$ is only clearly seen for angles $\theta \lesssim 15^{\circ}$ from the \emph{ab}-plane, making the role of inhomogeneous and/or surface superconductivity and related flux compression-like scenarios irrelevant to our experiments. For granular $d$-wave superconductors one could consider random-junctions \cite{sigrist}, which have also been claimed to cause the paramagnetic Meissner-effect. However, our sample is not by any means granular; it is in fact of extreme high quality as implied by the observation of the de Haas van Alphen-effect under fields well below 10 T. Paramagnetic impurities do not lead to such hysteresis nor ferromagnetism, as previously argued.

We might consider a complex scenario that involves a putative Fulde-Ferrel-Larkin-Ovchinnikov state as in CeCoIn$_5$ \cite{cecoin5}. However, the observation of this paramagnetic hysteretic signal over an extended region in temperatures, in sharp contrast to CeCoIn$_5$, points to an alternative scenario.

The high-field paramagnetic response can be reconciled with a superconducting state that breaks time reversal symmetry and carries intrinsic orbital angular momentum. The chiral states $k_z (k_x \pm  ik_y)$ proposed in Ref.~\onlinecite{kasahara} indeed possess orbital angular momentum along $\hat{z}$ direction. Nevertheless, if the angular momentum of this state was responsible for the paramagnetic response, it would be maximized for fields along the $\hat{z}$ axis. Since our observed paramagnetic response is restricted to a field orientation relatively close to the \emph{ab}-plane, either i) the original superconducting pairing symmetry is distinct from the proposed $k_z (k_x \pm  ik_y)$ wave function, or ii) a different, field-induced chiral paired state is realized when the field lies close to the \emph{ab}-plane, through mechanisms similar to the ones discussed in Ref. \onlinecite{li}. This raises the remarkable possibility of a field-induced transition to a hitherto unknown chiral state in URu$_2$Si$_2$.

\section{methods}
Single crystalline URu$_2$Si$_2$ was grown by the Czochralski method, electro-refined, and oriented by using a back-Laue CCD camera. Most results shown here were obtained from the same crystal previously used to study the evolution of the Fermi surface at very high fields \cite{altarawneh}  as well as the angular dependence of the upper-critical field at very low temperatures \cite{altarawneh2}. Torque measurements were performed by using a capacitive cantilever beam configuration. The angle relative to the field was measured with Hall probes.

\section{Acknowledgements}
We acknowledge discussions with N. Harrison. The NHMFL is supported by NSF through NSF-DMR-0084173 and the
State of Florida.  L.~B. is supported by DOE-BES through award DE-SC0002613.
The work at the LANL is carried out under the auspices of the U.S. DOE, Office of Science.

\section{Author contributions}

GL, BZ, QRZ, DR and LB performed the torque measurements and analyzed the data.
REB, PHT, FR, JDT, EDB synthesized, annealed and characterized the URu$_2$Si$_2$ single crystals.
PG provided theoretical modeling for the experimental observations.
LB and PG wrote the manuscript with the input of all co-authors
LB conceived the project.

\section{Competing Financial interests}
The authors declare no competing financial interests.

\end{document}